\documentclass[pra,onecolumn,preprint, 11pt]{revtex4}
\usepackage{amsmath}
\usepackage{graphicx}
\usepackage{amsfonts}

\begin{document}

\titlepage
\title{Two-Pulse Propagation in a Partially Phase-Coherent Medium}
\author{B.D. Clader and J.H. Eberly}
\affiliation{Department of Physics and Astronomy \\ University of Rochester \\ Rochester, NY 14627}

\begin{abstract}
We analyze the effects of partial coherence of ground state preparation on two-pulse propagation in a three-level $\Lambda$ medium, in contrast to previous treastments that have considered the cases of media whose ground states are characterized by probabilities (level populations) or by probability amplitudes (coherent pure states).  We present analytic solutions of the Maxwell-Bloch equations, and we extend our analysis with numerical solutions to the same equations.  We interpret these solutions in the bright/dark dressed state basis, and show that they describe a population transfer between the bright and dark state.  For mixed-state $\Lambda$ media with partial ground state phase coherence the dark state can never be fully populated.  This has implications for phase-coherent effects such as pulse matching, coherent population trapping, and electromagnetically induced transparency (EIT).  We show that for partially phase-coherent three-level media, self induced transparency (SIT) dominates EIT and our results suggest a corresponding three-level area theorem.
\end{abstract}

\maketitle

\section{Introduction}

The description of radiative phenomena in terms of intensities and probabilities is often satisfactory but the effects of wave coherences are then neglected. When coherence effects are prominent a wave description is usually adopted. Propagation of short laser pulses in resonant media often falls into an intermediate domain where neither description is satisfactory. The most important time scale is much longer than the period of the laser field but shorter than the decoherence time of the medium, and the slowly varying envelope and rotating wave approximations are engaged to simplify the description of evolution in this domain \cite{allen-eberly, shore, scully-zubairy}. 

However, theories of resonant propagation usually ignore the very real possibility that the state of the medium may be prepared in a way not adequately described by probability (population) assignments to the levels. The term ``phaseonium" was introduced by Scully  \cite{Scully} to describe a three-level atomic medium in the $\Lambda$ configuration, where two ground levels of the atoms are prepared in a phase-coherent pure-state superposition.  Phase coherent effects lead to coherent population trapping \cite{pop-trap, BDstate-early}, electromagnetically induced transparency (EIT) \cite{eit1,bib.Harris, eit-review}, pulse matching \cite{harris-matching1, harris-matching2, quantized-pulse-matching}, the dark area theorem \cite{Eberly-Kozlov},  simultons \cite{Konopnicki-Eberly}, and adiabatons \cite{Grobe-etal, Eberly-etal94}.  These effects are are all governed by dark-state \cite{Arimondo} considerations.  

Pure phaseonium is not easily prepared experimentally, and the question what effects may acompany more realistic preparation has remained open. Here we report new results obtained for propagation in media that could be called partial phaseonium, or ``mixonium", because the medium is allowed to have mixed state or partial coherence in its ground state. The simple cases we study are distinct from traditional EIT scenarios in that the ground state coherence is prepared ahead of any pulses entering the medium, while in traditional EIT it is the pump and probe pulses which prepare the medium coherence.  Methods to prepare a phaseonium medium have been shown for ladder \cite{pra-ladder-phaseonium} and $\Lambda$ \cite{kozlov-phaseonium} systems.

A previous study \cite{Kozlov-Eberly} by Kozlov and Eberly of two-pulse propagation in absorbing media made use of the Maxwell-Bloch model (see e.g. \cite{allen-eberly}), which describes fully coherent pulse propagation through atomic media. They reported two types of pulse evolution, one similar to that of one-pulse self-induced transparency (SIT) \cite{mccall-hahn}, and the other showing a different form of two-pulse evolution under conditions they identified with EIT, in which the dark state plays an important role. They defined the SIT-type propagation in three-level media as occurring when the dark state was nearly or completely empty and EIT-type propagation as occurring when the dark state was highly populated.  They showed that EIT-type dominated SIT-type for propagation in a fully coherent phaseonium medium.  However, when a $\Lambda$ medium is prepared without ground-state phase coherence, the role of the dark state is minimized, and we have shown  \cite{clader-eberly07, clader-eberly-pra07} that SIT-type propagation is dominant.

In realistic experimental preparation, pure-state three-level phaseonium is difficult to achieve. Ground state phase coherence can be lost due to environmental decoherence or imperfect preparation techniques.  Thus we report new results on two-pulse propagation through a $\Lambda$ medium that is prepared with the ground states in a partially phase-coherent superposition, between the two extremes of completely mixed or completely pure state, the basis for our term ``mixonium".  We include a variable parameter $\lambda$ in our initial state definition that will allow us to examine the propagation dynamics both analytically and numerically  over the entire range from completely mixed to completely pure.  We will follow the previous definition and determine whether it is EIT or SIT that dominates pulse propagation in mixonium.

We will use a three-regime language to distinguish three propagation zones, similar to our analysis of two-pulse propagation in a completely mixed-state medium \cite{clader-eberly-pra07}.  In the pure phaseonium case the analytic solutions to the Maxwel-Bloch (MB) equations predict that simulton pulses \cite{Konopnicki-Eberly} that begin entirely in the strongly interacting bright state will be transferred into the non-interacting dark state just as one would expect based on coherent population trapping and dark area theorem arguments.  If the medium is initially prepared only partially phase-coherent, a similar simulton transfer process occurs. However the lack of complete phase coherence prohibits the dark state from being fully populated, and the pulses will continue to interact with the ground to excited state transition.  Our analytic solutions allow us to determine the maximum population of the dark state, when the medium is not in a pure state.

The inability of the dark state to be fully populated, and the subsequent pulse-medium interaction that continues for non-pure medium preparation causes SIT-like effects and reduces the role of EIT.  This has dramatic consequences for the pulse matching that occurs in phaseonium.  We find that our analytic solutions have broad predictive capabilities for mixonium due to pulse-reshaping caused by SIT.  In the phaseonium case our numerical solutions agree with previous studies \cite{Kozlov-Eberly} predicting temporal pulse matching.  However, in the mixonium case SIT begins to play a role because the dark state cannot be fully populated.  We find that for long propagation distances SIT always dominates in a mixed-state medium. Further highlighting this effect, we also show that the bright pulse area behaves in a manner very similar to SIT, which leads us to believe that there is an equivalent three-level area theorem.

%%%%%%%%%%%%%%%%%%%%%%%%%%%%%%%%%%%%%%%%%%%%%%%%%
%
%						         Theoretical Model
%
%%%%%%%%%%%%%%%%%%%%%%%%%%%%%%%%%%%%%%%%%%%%%%%%%
\section{Physical Model}

\begin{figure}[h]
\begin{center}
\leavevmode
\includegraphics{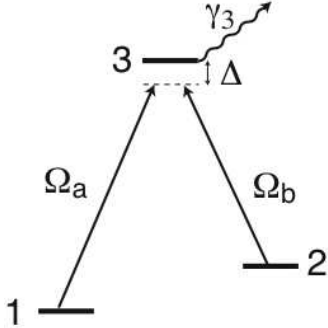}
\end{center}
\caption{\label{lambda-fig}Three level atom in the $\Lambda$ configuration with level $1$ connected to level $3$ via a laser field $\Omega_a$, which we refer to as the pump field, and level $2$ connected to level $3$ with laser field $\Omega_b$ which we refer to as the Stokes field.  We assume a two photon resonance condition so that both fields are detuned from resonance by an equal amount $\Delta$.  Loss from the excited state is included by a damping term $\gamma_3$.}
\end{figure}

We consider dual-pulse propagation in a medium of three-level atoms in the lambda configuration as shown in Fig. \ref{lambda-fig}.  Possible physical realizations appear to be in a D line of cesium or rubidium, and we give some of the experimental parameters at the end of this section. We assume a two-photon resonance condition such that both laser fields are detuned by an equal amount $\Delta$.  The Hamiltonian of the system in the rotating wave picture is given by
\begin{equation} \label{Hamiltonian}
H = \hbar\Delta |3\rangle\langle 3| -\hbar\frac{\Omega_{a}}{2}|1\rangle\langle 3| - 
\hbar\frac{\Omega_{b}}{2}|2\rangle\langle 3| 
- \hbar\frac{\Omega_{a}^*}{2}|3\rangle\langle 1| - 
\hbar\frac{\Omega_{b}^*}{2}|3\rangle\langle 2|,
\end{equation}
where $\Omega_a = 2\vec{d}_1\cdot \mathcal{\vec{E}}_a/\hbar$ is the Rabi frequency of the pump laser field.  Throughout this paper, we will refer to $\Omega_a$ as the pump field and $\Omega_b$ as the Stokes field to be consistent with the common nomenclature associated with stimulated Raman scattering.  Our notation implies that the laser fields can be represented as a slowly varying envelope times a carrier frequency, e.g., $\vec{E}_a = \mathcal{\vec{E}}_a e^{i(k_a x - \omega_a t)} + \rm{c.c.}$, where $k_a$ and $\omega_a$ are the wavenumber and carrier frequency of the pulse and $\mathcal{\vec{E}}_a$ is the envelope function.  The term $d_1$ is the dipole moment of the $1 \to 3$ transition.  Similar notation applies to the Stokes Rabi frequency $\Omega_b$.

Individual density matrix equations can be derived from the Hamiltonian and the von Neumann equation $i\hbar\partial{\rho}/\partial T = [H,\rho$].  They are given by:
\begin{subequations} 
\label{rhoEquations2}
\begin{align}
\frac{\partial\rho_{11}}{\partial T} &= i\frac{\Omega_a}{2}\rho_{31} - i\frac{\Omega_a^*}{2}\rho_{13} \\
\frac{\partial\rho_{22}}{\partial T} &= i\frac{\Omega_b}{2}\rho_{32} - i\frac{\Omega_b^*}{2}\rho_{23} \\
\frac{\partial\rho_{33}}{\partial T} &= -i\frac{\Omega_a}{2}\rho_{31} + i\frac{\Omega_a^*}{2}\rho_{13}
- i\frac{\Omega_b}{2}\rho_{32} + i\frac{\Omega_b^*}{2}\rho_{23} \\
\frac{\partial\rho_{12}}{\partial T} &=  i\frac{\Omega_a}{2}\rho_{32} -i\frac{\Omega_b^*}{2}\rho_{13} \\
\frac{\partial\rho_{13}}{\partial T} &= i\Delta \rho_{13} - i\frac{\Omega_b}{2}\rho_{12} + i\frac{\Omega_a}{2}(\rho_{33} - \rho_{11}) \\
\frac{\partial\rho_{23}}{\partial T} &=  i\Delta \rho_{23} - i\frac{\Omega_a}{2}\rho_{21} + i\frac{\Omega_b}{2}(\rho_{33} - \rho_{22}).
\end{align}
\end{subequations}
We assume that the temporal duration of each laser pulse envelope is sufficiently short that we can neglect decay terms, such as arising from loss to other atomic states, spontaneous emission, or collisional dephasing effects.  For alkali vapors this requires pulse durations on the order of, or shorter than, 1 ns.  By making use of the slowly varying envelope approximation and the rotating wave approximation we can reduce Maxwell's wave equation to two independent first order wave equations for each individual pulse.  They are given by:
\begin{subequations}
\label{MaxwellEquation2}
\begin{align}
\frac{\partial  \Omega_{a}}{\partial Z}& = - i\mu_a \int_{-\infty}^{\infty}d\Delta F(\Delta)\rho_{13} = -i\mu_a\langle\rho_{13}\rangle \\
\frac{\partial  \Omega_{b}}{\partial Z} & = - i\mu_b \int_{-\infty}^{\infty}d\Delta F(\Delta)\rho_{23} = -i\mu_b\langle\rho_{23}\rangle,
\end{align}
\end{subequations}
where we have written equations \eqref{rhoEquations2} and \eqref{MaxwellEquation2} in a retarded time coordinate system such that $T = t-x/c$ and $Z=x/c$.  Thus the derivatives are given by: $\partial /\partial t = \partial /\partial T$ and $\partial /\partial Z = c\partial /\partial x + \partial /\partial t$.  

We use the bracket notation to symbolize a statistical average to take into account inhomogeneous broadening, for example due to thermal motion of the atoms if the medium is a vapor. When needed, the average is performed with the function $F(\Delta) = T_2^*/\sqrt{2\pi}e^{-(T_2^*)^2(\Delta-\bar\Delta)^2/2}$, where $T_2^*$ is the inhomogeneous lifetime and $\bar\Delta$ is the detuning of the laser fields from atomic line center. We will consistently assume line-center tuning, so $\bar\Delta = 0$ is implied throughout.  The parameters $\mu_a = N d_1^2\omega_a/\hbar \epsilon_0$ and $\mu_b = N d_2^2\omega_b/\hbar \epsilon_0$, where $N$ is the density of the atoms, are proportional to the usual attenuation coefficient or inverse Beer's length:
\begin{equation} \label{InvBeerCoeff}
\alpha_D(\Delta) = \pi F(\Delta) \mu/c \quad \to \quad \alpha_D(0) = \sqrt{\pi/2}T_2^*\mu/c,
\end{equation}
for each transition, where the simplified final form applies to line-center tuning.  When $\mu_a = \mu_b \equiv \mu$, which we will assume hereafter, Eqs. \eqref{rhoEquations2} and \eqref{MaxwellEquation2} are exactly solvable by using methods such as inverse scattering \cite{early-IS, akns-IS} or B\"acklund transformations \cite{BacklundBook, Lamb-BTreview, park-shin}.

A single weak pulse presents a familiar case, in which the pulse causes some population exchange between ground and excited states, and even though we neglect homogeneous loss terms, Doppler (or any other inhomogeneous) broadening serves as a dephasing mechanism.  Thus a single weak pulse will be absorbed as it propagates, and its peak intensity will decay exponentially as $|\Omega(x)|^2 = |\Omega(0)|^2 e^{-\alpha_D x}$, where $\Omega(0)$ is the peak intensity of the weak input pulse and $\alpha_D^{-1}$ is the Doppler absorption depth or Beer's length given in (\ref{InvBeerCoeff}).  Two separate mechanisms can change these familiar absorptive properties.  These are self-induced transparency (SIT) which arises from dynamic nonlinearities associated with a strong and coherent pulse, and electromagnetically induced transparency (EIT) which arises when a second laser pulse interacts with the same excited state, thus inducing two-photon coherence between two different ground states. 

Potential physical realizations of such a system include the D line transitions for Rubidium or Cesium.  For the $D_2$ line of Rubidium, which we use as our model in this paper, the excited state lifetime is around $30$ ns and the Doppler lifetime for room temperature vapor is about $0.5$ ns, while the Beer's length is approximately $1$ cm.  We assume that the laser pulses are linearly polarized with a bandwidth greater than the hyperfine splittings of the excited state, which is approximately $500$ MHz, but with bandwidth narrow enough to resolve the two hyperfine ground states which are split by approximately $10$ GHz.  This implies the effective dipole moments, $d_1$ and $d_2$, are equal which means that taking $\mu_a = \mu_b$ is quite accurate. Our calculations use pulses with temporal duration around $2.5$ ns, which is consistent with all of our approximations and assumptions.

%%%%%%%%%%%%%%%%%%%%%%%%%%%%%%%%%%%%%%%%%%%%%%%%%
%
%						Bright Dark States
%
%%%%%%%%%%%%%%%%%%%%%%%%%%%%%%%%%%%%%%%%%%%%%%%%%
\section{Bright - Dark States}\label{ss:bright-dark}
The three-level MB model permits a useful parameterization in terms of bright and dark states \cite{Arimondo}, which help explain the interference effects caused by ground state coherence.  We define the Bright and Dark states as:
\begin{equation} \label{BrightDarkDefs3}
|B\rangle \equiv \frac{1}{\Omega_T}\left(\Omega_{a}|1\rangle + \Omega_{b}|2\rangle\right) \quad {\rm and}\quad |D\rangle \equiv \frac{1}{\Omega_T}\left(\Omega_{b}^*|1\rangle - \Omega_{a}^*|2\rangle\right),
\end{equation}
where $\Omega_T = (|\Omega_a|^2 + |\Omega_b|^2)^{1/2}$ is the ``total" Rabi frequency.  In terms of the bright and dark states the Hamiltonian in Eq. \eqref{Hamiltonian} can be written simply as:
\begin{equation}
\label{BDHam}
H = \hbar\Delta |3\rangle\langle 3| -\frac{\Omega_T}{2}|B\rangle\langle 3| 
- \frac{\Omega_T}{2}|3\rangle \langle B|.
\end{equation}
The interaction terms in $H$ clearly depend only on $|B\rangle$.  The orthogonal dark state $|D\rangle$ does not participate in the temporal dynamics so once population enters the dark state it becomes trapped, unless it evolves as a result of propagation.  As clearly seen in the definition, the bright and dark state basis is a fully coherent, pure-state superposition of the two ground states.  This basis helps provide a clear understanding of phase-coherent effects on two-pulse propagation.  

To more clearly understand these effects it is useful to convert the MB model into the bright-dark basis.  This was done by Fleischhauer and Manka \cite{BDstates}, and we will follow their lead.  We will take a pure-state approach in this particular section, to focus on purely phase-coherent effects.  Assuming a pure-state wavefunction of the form $|\psi\rangle = c_1|1\rangle + c_2|2\rangle + c_3|3\rangle$, the probability amplitude equations in the original atomic basis are:
\begin{subequations}
\label{mix-Schrodinger}
\begin{align}
\dot{c}_1 &= i\frac{\Omega_a}{2}c_3 \\
\dot{c}_2 &= i\frac{\Omega_b}{2}c_3 \\
\dot{c}_3 &= i\frac{\Omega_a^*}{2}c_1 + i\frac{\Omega_b^*}{2}c_2 -i\Delta c_3,
\end{align}
\end{subequations}
where the dot refers to $\partial/\partial T$.  One can recover the density matrix given in Eqs. \eqref{rhoEquations2} by taking $\rho = | \psi \rangle\langle\psi |$.  From Eq. \eqref{BrightDarkDefs3} we can calculate the bright and dark probability amplitudes in terms of $c_1$ and $c_2$.  They are:
\begin{subequations}
\label{BDprobamp}
\begin{align}
c_B & = \frac{1}{\Omega_T}(\Omega_a^* c_1 + \Omega_b^* c_2) \\
c_D & = \frac{1}{\Omega_T}(\Omega_b c_1 - \Omega_a c_2).
\end{align}
\end{subequations}
For simplicity we will assume that the fields are unchirped and real, giving $\Omega_a = \Omega_a^*$ and $\Omega_b = \Omega_b^*$.  This allows us to rewrite the Eqs. \eqref{mix-Schrodinger} as:
\begin{subequations}
\label{BDSchrodinger}
\begin{align}
\dot{c}_B &= i\frac{\Omega_D}{2} c_D + i\frac{\Omega_T}{2} c_3 \\
\dot{c}_D &= - i\frac{\Omega_D}{2} c_B \\
\dot{c}_3 &= i\frac{\Omega_T}{2}c_B - i\Delta c_3,
\end{align}
\end{subequations}
where
 \begin{equation}
 \label{DRabiFreq}
 \Omega_D = \frac{2i}{\Omega_T^2}(\Omega_a\dot{\Omega}_b - \Omega_b\dot{\Omega}_a)
 \end{equation}
 is called the the dark Rabi frequency.  We can write the corresponding Maxwell's equations for the total and dark Rabi frequencies as:
\begin{subequations}
\label{BDMaxwell}
\begin{align}
\frac{\partial \Omega_T}{\partial Z} & = -i\mu \langle c_B c_3^*\rangle \\
\frac{\partial \Omega_D}{\partial Z} & = -2\mu\frac{\partial}{\partial T}\left\langle \frac{c_D c_3^*}{\Omega_T}\right\rangle,
\end{align}
\end{subequations}
where $\mu_a = \mu_b \equiv \mu$ as previously noted.

A surprising result occurs if the pulses are matched temporally, i.e. $\dot\Omega_a/{\Omega}_a = \dot\Omega_b/{\Omega}_b$, which gives $\Omega_D = 0$.  This allows us to write Eqs. \eqref{BDSchrodinger} and \eqref{BDMaxwell} as:
\begin{subequations}
\label{BDtwolevel}
\begin{align}
\dot c_B & = i\frac{\Omega_T}{2}c_3 \\
\dot c_3 & = i\frac{\Omega_T}{2}c_B - i\Delta c_3 
\end{align}
\end{subequations}
and
\begin{equation}
\label{BMaxwelltwolevel}
\frac{\partial\Omega_T}{\partial Z} = -i\mu\langle c_B c_3^*\rangle.
\end{equation}
We see that all dark-state effects disappear, and Eqs. \eqref{BDtwolevel} and \eqref{BMaxwelltwolevel} are identical in form to the Maxwell-Bloch equations for a single ``total" pulse interacting with a two-level atom.  Thus if the dark state is initially unpopulated and the pulses are matched temporally, then the population is constrained to states $|B\rangle$ and $|3\rangle$, and the three-level behavior is identical to that of a two-level atom.  This explains the three-level simulton solutions derivation \cite{Konopnicki-Eberly}.  However we will find that small fluctuations prohibit such isolation of the dark state over long propagation distances.

Kozlov and Eberly (KE) also used this fact to examine SIT-type vs. EIT-type transparency \cite{Kozlov-Eberly}.  In SIT it is nonlinearities that cause the medium to be transparent even while the dark state is unpopulated and strong interaction occurs between the pulse and the medium.  Thus SIT propagation occurs when $|c_D|^2 \approx 0$.  As just shown, this implies that the three-level equations can be reduce to two-level form, and transparency only occurs if the two pulses are matched.  In EIT it is the presence of a second pulse and population trapping in the dark state that cancels absorption of the pulses by the medium.  Thus KE defined EIT propagation to occur when $|c_D|^2 \approx 1$.  In this case, the medium is transparent because the combined pulse-medium system is in the dark state, cancelling pulse-medium interaction.  We will use this same definition throughout this paper.

%%%%%%%%%%%%%%%%%%%%%%%%%%%%%%%%%%%%%%%%%%%%%%%%%
%
%						         Mixonium Analytical Solutions
%
%%%%%%%%%%%%%%%%%%%%%%%%%%%%%%%%%%%%%%%%%%%%%%%%%
\section{Mixonium Analytical Solutions}
We wish to analyze two-pulse propagation through a partially coherent $\Lambda$ system, i.e. a medium prepared with $\rho_{ij} \ne 0$ but $|\rho_{ij}|^2 < \rho_{11}\rho_{22}$ for $i \ne j$.  We do this by solving Eqs. \eqref{rhoEquations2} and \eqref{MaxwellEquation2} by using the Park-Shin (PS) B\"acklund method \cite{park-shin}.  We take the initial density matrix of each atom to be in a partially-coherent superposition of the two ground states, which we write in explicit form, for real $\alpha$ and $\beta$, as
\begin{equation}
\label{mix-InCondition}
\rho^{(0)} = \begin{pmatrix}
\alpha^2 & \lambda\alpha\beta e^{i\phi} & 0 \\
\lambda \alpha\beta e^{-i \phi} & \beta^2 & 0 \\
0 & 0 & 0
\end{pmatrix},
\end{equation}
where $\alpha^2$ and $\beta^2$ are the populations of ground states 1 and 2 respectively (we will always take the case $\alpha^2 > \beta^2$), and $\phi$ is the phase of the partial coherence.  We introduce a coherence parameter $\lambda$ that takes values between 0 and 1, between a completely mixed state and a pure state respectively.  We will assume the fields are tuned to the center of the inhomogeneous line for the two transitions.  To find the solution, first we diagonlize $\rho^{(0)}$ in Eq. \eqref{mix-InCondition} with the rotation matrix
\begin{equation}
\label{S_trans}
S=
\begin{pmatrix}
\cos\theta & \sin \theta e^{i\phi} & 0 \\
-\sin\theta e^{-i\phi} & \cos \theta & 0 \\
0 & 0 & 1
\end{pmatrix},
\end{equation}
where
\begin{subequations}
\label{sin-cos-def}
\begin{align}
\cos \theta & = \frac{\zeta - \beta^2}{\sqrt{(\zeta-\beta^2)^2  + \lambda^2 \alpha^2 \beta^2}} \\
\sin \theta & = \frac{-\lambda \alpha \beta}{\sqrt{(\zeta-\beta^2)^2  + \lambda^2 \alpha^2 \beta^2}},
\end{align}
\end{subequations}
and
\begin{equation}
\label{eigValue}
\zeta=\frac{1}{2}\{1 + [1-4(1-\lambda^2)\alpha^2\beta^2]^{1/2}\},
\end{equation}
where $\zeta$ and 1-$\zeta$ are the positive non-zero eigenvalues of the matrix $\rho^{(0)}$.  The rotation matrix $S$ has an additional degree of freedom in that the first and second columns can be multiplied by an arbitrary phase, while still diagonalizing $\rho^{(0)}$.  In some plots of our analytic solutions we will make use of this fact.

One can verify that $S$ diagonalizes $\rho^{(0)}$ in Eq. \eqref{mix-InCondition} through the operation
\begin{equation}
\label{DiagInCondition}
\rho^{(0)}_{\text{d}} = S^\dag \rho^{(0)} S =  \begin{pmatrix}
\zeta & 0 & 0 \\
0 & 1-\zeta & 0 \\
0 & 0 & 0
\end{pmatrix}.
\end{equation}

We can solve the much simpler problem of pulses propagating through a medium prepared with initial density matrix given by Eq. \eqref{DiagInCondition} by using the Park-Shin \cite{park-shin} B\"acklund transformation technique, just as we did in a previous paper \cite{clader-eberly-pra07}.  These pulse solutions are:
\begin{subequations}
\label{DiagPulseSol}
\begin{align}
\label{om1_d} \Omega_a^{\text{(d)}} & = \frac{4}{\tau}\left[2\cosh\left(\frac{T}{\tau}-\zeta\kappa Z \right)+\text{exp}\left(\frac{T}{\tau}+\kappa Z (3\zeta-2)\right)\right]^{-1} \\
\label{om2_d} \Omega_b^{\text{(d)}} & = \frac{4}{\tau}\left[2\cosh\left(\frac{T}{\tau}-\kappa Z(1-\zeta)\right)+\text{exp}\left(\frac{T}{\tau}+\kappa Z(1-3\zeta)\right)\right]^{-1}.
\end{align}
\end{subequations}
where $\tau$ is the nominal pulse width, the superscript (d) is to remind us that these are the solutions for two pulses propagating through a medium with a diagonal density matrix, and $\kappa/c$ is the inverse absorption length given by:
\begin{equation}
\label{mix-lengthScale}
\kappa =  \frac{\mu\tau}{2} \int_{-\infty}^{\infty}\frac{F(\Delta)d\Delta}{1+(\Delta\tau)^2}.
\end{equation}
In the limit where $T_2^* \ll  \tau$, the scaled inverse absorption depth $2\kappa/c$ becomes the inverse Doppler absorption depth such that $2\kappa / c \to \alpha_D$, which we previously defined as $\alpha_D = \sqrt{\pi/2} \mu T_2^*/c$.

The solutions to the density matrix elements for a particular value of the detuning $\Delta$ are given by:
\begin{subequations}
\label{DiagDensityMatrixSolution}
\begin{align}
\rho_{11}^{(\text{d})} & = \frac{1}{1+(\Delta\tau)^2}\bigg\{\zeta[|f_{11}|^2+(\Delta \tau)^2] + (1-\zeta)|f_{12}|^2\bigg\} \label{mix-ground1} \\
\rho_{22}^{(\text{d})} & = \frac{1}{1+(\Delta\tau)^2}\bigg\{\zeta|f_{12}|^2 + (1-\zeta)[|f_{22}|^2+(\Delta\tau)^2]\bigg\} \\
\rho_{33}^{(\text{d})} & = \frac{1}{1+(\Delta\tau)^2}\bigg(\zeta|f_{13}|^2 + (1-\zeta)|f_{23}|^2\bigg) \label{mix-excited_state}  \\
\rho_{12}^{(\text{d})} & = \frac{1}{1+(\Delta\tau)^2}\bigg[\zeta(f_{11}-i\Delta\tau)f_{12} + (1-\zeta)(f_{22}+i\Delta\tau)f_{12}\bigg] \\
\label{mix-Coherence13} \rho_{13}^{(\text{d})} & = \frac{1}{1+(\Delta\tau)^2}\bigg[\zeta(f_{11}-i\Delta\tau)f_{13} + (1-\zeta)f_{12}f_{23}\bigg] \\
\label{mix-Coherence23} \rho_{23}^{(\text{d})} & =\frac{1}{1+(\Delta\tau)^2}\bigg[\zeta f_{12}^*f_{13} + (1-\zeta)(f_{22}-i\Delta\tau)f_{23}\bigg],
\end{align}
\end{subequations}
where the functions $f_{ij} $ are:
\begin{subequations}
\label{mix-fFunctions}
\begin{align}
f_{11} & = \bigg\{2 \textnormal{ sinh} \big(T/\tau-\zeta \kappa Z\big) - 
\exp\big[T/\tau+(3\zeta-2)\kappa Z \big]\bigg\}\bigg/D(Z,T) \\
%\\
f_{22} & = \bigg\{-2 \textnormal{ cosh }\big(T/\tau-\zeta \kappa Z\big) + 
\exp\big[T/\tau+(3\zeta-2)\kappa Z\big]\bigg\}\bigg/D(Z,T) \\
%\\
f_{12} & = 2 e^{T/\tau-(1-\zeta) \kappa Z}/D(Z,T) \\
%\\
f_{13} & = 2 i/D(Z,T) \\
%\\
f_{23} & = 2 i e^{(2\zeta-1) \kappa Z}/D(Z,T),
\end{align}
\end{subequations}
and the denominator function $D(Z,T)$ is given by:
\begin{subequations}
\begin{align}
\label{mix-DFunction}
D(Z,T) & = 2 \textnormal{ cosh }\big(T/\tau-\zeta \kappa Z\big) + \exp\big[T/\tau+(3\zeta-2) \kappa Z\big].
\end{align}
\end{subequations}

Now that we have the solutions for a medium initially in the diagonal state, we obtain the mixonium solutions through the operations
\begin{equation}
\label{mix-PulseSolution}
\begin{pmatrix}
\Omega_a \\
\Omega_b
\end{pmatrix}
=
\begin{pmatrix}
\cos\theta & \sin\theta e^{i\phi} \\
-\sin\theta e^{-i\phi} & \cos\theta
\end{pmatrix}
\begin{pmatrix}
\Omega_a^{\text{(d)}} \\
\Omega_b^{\text{(d)}}
\end{pmatrix},
\end{equation}
and
\begin{equation}
\label{mix-DensityMatrixSolution}
\rho=S \rho^{\text{(d)}} S^\dag =
\begin{pmatrix}
\cos\theta & \sin\theta e^{i\phi} & 0 \\
-\sin\theta e^{-i\phi} & \cos \theta& 0 \\
0 & 0 & 1
\end{pmatrix}
\begin{pmatrix}
\rho_{11}^{(d)} & \rho_{12}^{(d)} & \rho_{13}^{(d)} \\
\rho_{21}^{(d)} & \rho_{22}^{(d)} & \rho_{23}^{(d)} \\
\rho_{31}^{(d)} & \rho_{32}^{(d)} & \rho_{33}^{(d)}
\end{pmatrix}
\begin{pmatrix}
\cos\theta & -\sin\theta e^{i\phi} & 0 \\
\sin\theta e^{-i\phi} & \cos \theta& 0 \\
0 & 0 & 1
\end{pmatrix}
\end{equation}
Eqns. \eqref{mix-PulseSolution} and \eqref{mix-DensityMatrixSolution} are the exact solutions to Eqns. \eqref{rhoEquations2} and \eqref{MaxwellEquation2} for a medium initially prepared in a mixed-state coherent superposition of the two ground states, as given in Eq. \eqref{mix-InCondition}.  One can see that the pulse and density matrix solutions are simply a rotation of the completely mixed-state solutions that we have previously solved \cite{clader-eberly-pra07}, with the eigenvalue of the initial density matrix taking the place of the ground state population.  The invariance of the MB equations under such operations, which allows us to obtain the mixonium solutions, is discussed in further detail by Park and Shin \cite{park-shin}.  It also complements previous numerical work which demonstrated the applicability of such  ``dressed-field" pulses \cite{Eberly-etal94}.

%%%%%%%%%%%%%%%%%%%%%%%%%%%%%%%%%%%%%%%%%%%%%%%%%
%
%						         Pure-State Analysis
%
%%%%%%%%%%%%%%%%%%%%%%%%%%%%%%%%%%%%%%%%%%%%%%%%%
\section{Phaseonium Analytical Solution Analysis}
The pulse and density matrix solutions presented are clearly quite complicated.  However, we can begin to understand them by starting with the pure state case where $\lambda=1$.  In that case Eq. \eqref{eigValue} becomes $\zeta = 1$ and Eq. \eqref{sin-cos-def} simplifies to $\cos\theta=\alpha$ and $\sin\theta = -\beta$.  The pulse solutions then simplify to:
\begin{subequations}
\label{PureStatePulse}
\begin{align}
\Omega_a & = \alpha\Omega_a^{\text{(d)}} - \beta e^{i\phi}\Omega_b^{\text{(d)}} \\
\Omega_b & = \beta e^{-i \phi} \Omega_a^{\text{(d)}} + \alpha\Omega_b^{\text{(d)}}.
\end{align}
\end{subequations}
The atomic solutions also simplify substantially.  Since we are considering pure states, we can consider just the wavefunction.  Thus Eq. \eqref{mix-DensityMatrixSolution} can be simplified to:
\begin{equation}
\label{mix-probAmpSolution}
|\psi\rangle=S |\psi^{\text{(d)}}\rangle=
\begin{pmatrix}
\alpha & -\beta e^{i\phi} & 0 \\
\beta e^{-i\phi} & \alpha & 0 \\
0 & 0 & 1
\end{pmatrix}
\begin{pmatrix}
c_{1}^{(d)} \\
c_{2}^{(d)} \\
c_{3}^{(d)}
\end{pmatrix},
\end{equation}
where the diagonal probability amplitude solutions are given by:
\begin{subequations}
\label{DiagProbAmpSolution}
\begin{align}
c_{1}^{(\text{d})} & = \frac{1}{\sqrt{1+(\Delta\tau)^2}}[f_{11}-i \Delta \tau] \\
c_{2}^{(\text{d})} & = \frac{1}{\sqrt{1+(\Delta\tau)^2}}f_{12} \\
c_{3}^{(\text{d})} & = \frac{1}{\sqrt{1+(\Delta\tau)^2}}f_{13}^*,
\end{align}
\end{subequations}
which can be derived from Eqs. \eqref{mix-ground1} - \eqref{mix-excited_state} up to an overall phase.  Thus the probability amplitudes on line-center (taking $\Delta = 0$) can be written in a substantially simplified form as:
\begin{subequations}
\label{PureStateAmplitude}
\begin{align}
c_1 & = \alpha f_{11} - \beta f_{12} \\ 
c_2 & = \beta f_{11} + \alpha f_{12} \\
c_3 & = f_{13}^*
\end{align}
\end{subequations}
where we have assumed the phase of the ground state coherence $\phi = 0$ for simplicity, and we will do so for the remainder of this paper.  Eqns. \eqref{PureStatePulse} and \eqref{PureStateAmplitude} mark a substantial simplification of the general mixed state solution. 

\subsection{Phaseonium Input Regime: $-\kappa Z \gg 1$}
We can further simplify the analysis by dividing the evolution into three distinct regimes.  We will consider the asymptotic ``input" as regime I, the asymptotic ``output" as regime III, and the transfer zone in between as regime II.  We study regime I by taking $-\kappa Z \gg 1$.  In this limit the pulse solutions become:
\begin{subequations}
\label{PulseInput}
\begin{align}
\Omega_a & \to \alpha\frac{2}{\tau}\text{ sech} \bigg(\frac{T}{\tau} - \kappa Z\bigg) \\
\Omega_b & \to \beta\frac{2}{\tau}\text{ sech} \bigg(\frac{T}{\tau} - \kappa Z\bigg),
\end{align}
\end{subequations}
and the line-center probability amplitudes are given by
\begin{subequations}
\label{AmplitudeInput}
\begin{align}
c_1 & \to \alpha\tanh\bigg(\frac{T}{\tau} - \kappa Z\bigg) \\
c_2 & \to \beta\tanh\bigg(\frac{T}{\tau} - \kappa Z\bigg) \\
c_3 & \to - i \text{ sech}\bigg(\frac{T}{\tau} - \kappa Z\bigg).
\end{align}
\end{subequations}
These pulse and amplitude solutions are exactly the matched simulton solutions of Konopnicki and Eberly \cite{Konopnicki-Eberly} moving with group velocity $v_g / c = (1+\kappa \tau)^{-1}$.  This is an example of SIT type propagation, because $|c_D|^2 = 0$ in the limit, and the excited state is fully populated at the pulse peak. However because of the particular pulse shape and their matching, the medium is transparent, but with a slowed velocity, possibly much slower than $c$.

\subsection{Phaseonium Output Regime: $\kappa Z \gg 1$}
Similarly the output regime III can be considered by taking $\kappa Z \gg 1$.  In this limit the pulse solutions are:
\begin{subequations}
\label{PulseOutput}
\begin{align}
\Omega_a & \to -\beta\frac{2}{\tau}\text{ sech} \bigg(\frac{T}{\tau} \bigg) \\
\Omega_b & \to \alpha\frac{2}{\tau}\text{ sech} \bigg(\frac{T}{\tau} \bigg) ,
\end{align}
\end{subequations}
and the line-center probability amplitudes are simply the constant values
\begin{subequations}
\label{AmplitudeOutput}
\begin{align}
c_1 & \to -\alpha \\
c_2 & \to -\beta \\
c_3 & \to 0.
\end{align}
\end{subequations}
Just as in the input regime I, these pulse solutions are also matched simultons but now moving with the vacuum velocity $c$.  One can see from Eqns. \eqref{AmplitudeOutput} that the excited state probability amplitude is 0.  This is because all population is now in the dark state, such that $|c_D|^2 = 1$ in the limit.  Thus the ground state amplitudes become constant and EIT type propagation occurs, allowing the pulses to propagate without any interaction with the medium.  We note from Eqs. \eqref{PulseOutput} that the pulse shapes are matched, and with ratio
\begin{equation}
\label{pulse-ratio}
\frac{\Omega_a}{\Omega_b} = -\frac{\beta}{\alpha},
\end{equation}
which is exactly the ratio predicted by the Dark Area Theorem \cite{Eberly-Kozlov}.

As remarked previously when discussing the Hamiltonian, we know that the dark state is completely decoupled from the dynamics.  Thus if during the propagation the pulses become matched to the medium, as in (\ref{pulse-ratio}), no further population transfer can occur.  We see that when $\lambda=1$, meaning the initial state of the density matrix is a pure state, the analytic solutions describe a transfer of SIT-type simulton pulses to EIT-type pulses in the dark state, where the population is trapped and remains constant.  We also see from these analytic solutions, that the original simulton solutions \cite{Konopnicki-Eberly} are simply limiting cases of our more general solutions.  The simulton solutions propagate in stable form without any change and without ever becoming trapped in the dark state.  However this picture is not complete as we now see that transfer to the dark state still occurs.

\begin{figure}[h!]
\begin{center}
\leavevmode
\includegraphics[height=2.6in]{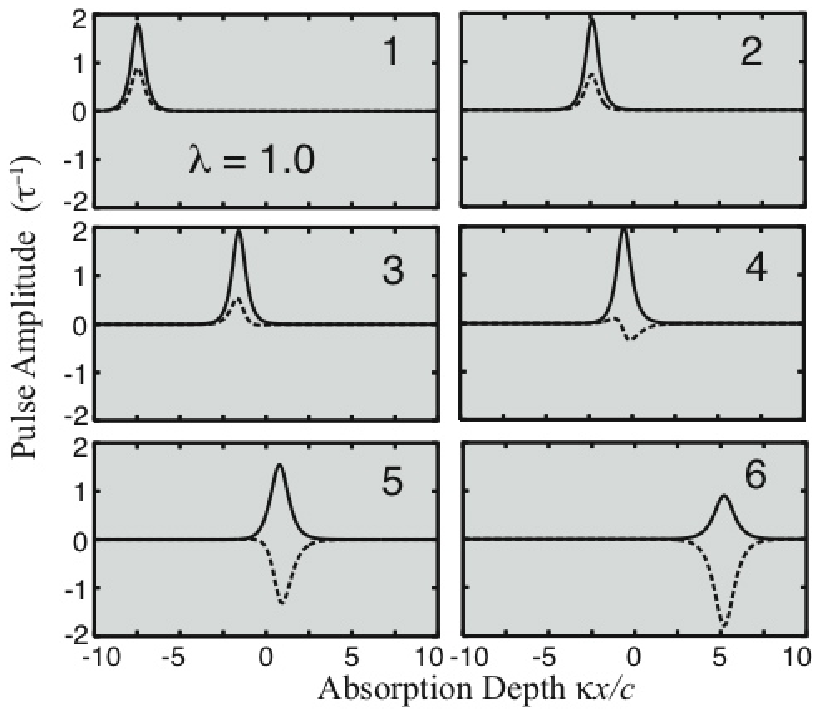}
\hspace{0.2in}
\includegraphics[height=2.3in]{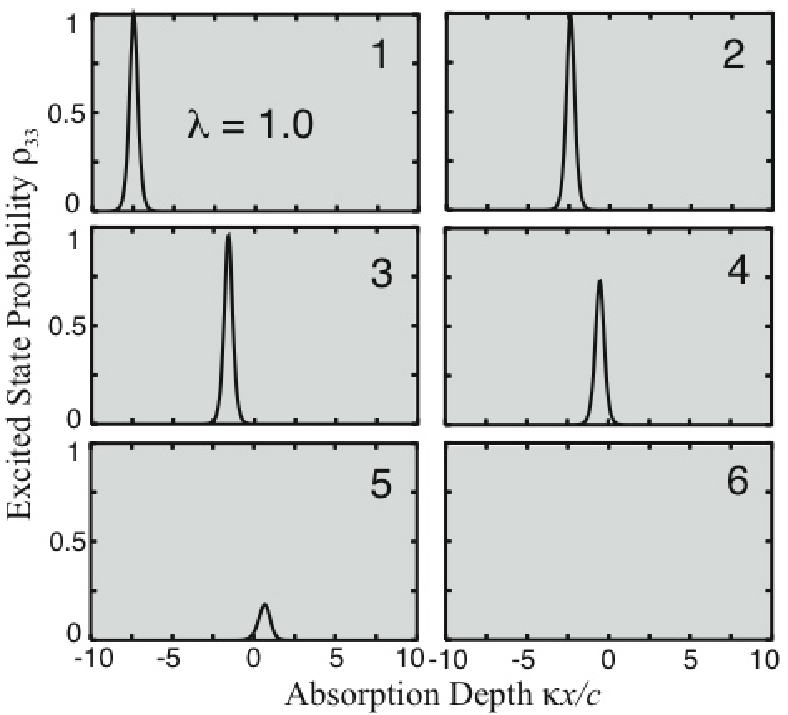}
\end{center}
\caption[Plots of the analytic pulse solutions given in Eq. \eqref{PureStatePulse}.]{\label{fig.mix.anPulse1.0}\label{fig.mix.anPop1.0} Plots of the analytic pulse solutions given in Eq. \eqref{PureStatePulse} on the left, and of the analytic excited state population solutions given in Eq. \eqref{PureStateAmplitude} on the right.  The pump pulse, $\Omega_a$ is the solid curve, and the Stokes pulse, $\Omega_b$ is the dashed curve.  The horizontal axis is $x$ in units of $\kappa/c$. The vertical axis is the pulse Rabi frequency in units of $\tau^{-1}$ (left) and the excited state probability $\rho_{33} = |c_3|^2$ (right). The background is slightly shaded to indicate the presence of the lambda medium.  The solid curve is the pump pulse, $\Omega_{a}$, and the dashed curve is the Stokes pulse, $\Omega_{b}$.  The plot shows the simulton transfer process as an exchange process between bright-state simultons to dark-state simultons. The excited state is heavily populated during the input propagation regime, when the pulses and atoms are in the bright state.  However as the pulses and atoms transfer into the dark state, the atoms no longer absorb any pulse energy, and the medium becomes transparent.  Parameters: $\alpha^2 = 0.8$, $\beta^2 = 0.2$, $\tau = 3T_2^*$, and $\lambda=1.0$.}
\end{figure}

We plot the analytic pulse solutions, $\Omega_a$ and $\Omega_b$, given in Eqs. \eqref{PureStatePulse}, in the left frame of Fig. \ref{fig.mix.anPulse1.0}, and the line-center excited state probability, $|c_3|^2$, given in Eq. \eqref{PureStateAmplitude}, in the right frame of Fig. \ref{fig.mix.anPop1.0}.  The numbers in each frame correspond to a particular time point (note that the time points are chosen to illustrate interesting changes that are occurring and are not uniformly spaced).  Examining both of these figures together we clearly see that in the initial simulton regime I, the pulses propagate in the bright state, causing excitation into the excited state (frames 1 and 2).  Then in the transfer regime II (frames 3-5), we see the relative magnitude of the pulses changing, and the phase of the Stokes pulse changing sign, along with a decrease in the excited state probability.  Finally, in regime III (frame 6) we see both pulses propagating without change and the upper level is not excited at all, since the pulses and medium are now in the dark state.

%%%%%%%%%%%%%%%%%%%%%%%%%%%%%%%%%%%%%%%%%%%%%%%%%
%
%						        Mixed State Analysis
%
%%%%%%%%%%%%%%%%%%%%%%%%%%%%%%%%%%%%%%%%%%%%%%%%%
\section{Mixed State Analysis}\label{ss:mixed-state-an}
We now consider the pulse and density matrix solutions for a medium prepared in an arbitrary mixed initial state (i.e., $0 < \lambda < 1$).  From Eq. \eqref{mix-PulseSolution}, we obtain the individual pulse solutions which are given by:
\begin{subequations}
\label{MixedStateSolutions}
\begin{align}
\Omega_a & = \cos \theta \Omega_a^{\text{(d)}} + \sin\theta \Omega_b^{\text{(d)}} \\
\Omega_b & = -\sin\theta \Omega_a^{\text{(d)}} + \cos\theta \Omega_b^{\text{(d)}},
\end{align}
\end{subequations}
where $\cos\theta$ and $\sin\theta$ are given in Eq. \eqref{sin-cos-def} and $\Omega_a^{(d)}$ and $\Omega_b^{(d)}$ are defined in Eq. \eqref{DiagPulseSol}.  The mixed-state density matrix solutions are given in Eq. \eqref{mix-DensityMatrixSolution} and are clearly quite cumbersome for this general mixed state case.  The simplest of these, and the one which provides the most insight into the physics is the excited state probability, $\rho_{33}$.  Written explicitly the line-center solution is:
\begin{equation}
\label{ExStateProb}
\rho_{33} =  \frac{ 4\big[\zeta + (1-\zeta) e^{(2\zeta-1)\kappa Z}\big]}{\big[2 \text{ cosh}\big(T/\tau - \zeta \kappa Z\big) + \exp\big(T/\tau + (3\zeta -2)\kappa Z\big)\big]^2}.
\end{equation}
Once again we will examine these solutions in the input and output regimes to help understand their underlying features.

\subsection{Mixonium Input Regime: $-\kappa Z \gg 1$}
In the input regime I, by taking the limit $-\kappa Z \gg 1$,  we find:
\begin{subequations}
\label{MixedPulseInput}
\begin{align}
\Omega_a & \to \cos\theta\frac{2}{\tau}\text{ sech} \bigg(\frac{T}{\tau} - \zeta \kappa Z\bigg) \\
\Omega_b & \to -\sin\theta \frac{2}{\tau}\text{ sech} \bigg(\frac{T}{\tau} - \zeta \kappa Z\bigg),
\end{align}
\end{subequations}
while the excited state probability is:
\begin{equation}
\label{ExcStateProbInput}
\rho_{33} \to \zeta\text{ sech}^2\bigg(\frac{T}{\tau}-\zeta \kappa Z\bigg).
\end{equation}
Just as in the pure state case, Eqns. \eqref{MixedPulseInput} are matched simulton pulses.  However they are now generalized to non-pure states.  The most immediate differences between the two cases are the modification to the pulse amplitudes, and that the excited state no longer reaches a value of $\rho_{33} = 1$ at the pulse peak, but rather $\rho_{33} = \zeta$.  This results in a  group velocity of the pulses, $v_g / c= (1 + \zeta\kappa \tau)^{-1}$, which is different by the presence of the factor $\zeta$ relative to the pure state case.  This indicates that the interaction between the pulses and the medium is directly affected by the medium's mixed-state nature, and that the interaction strength is governed by the parameter $\zeta$, which is simply the eigenvalue of the initial mixonium density matrix.

\subsection{Mixonium Output Regime: $\kappa Z \gg 1$}
We look at the mixed-state output regime III by taking $\kappa Z \gg 1$.  In this limit the pulse solutions are:
\begin{subequations}
\label{MixedPulseOutput}
\begin{align}
\Omega_a & \to \sin\theta\frac{2}{\tau}\text{ sech} \bigg(\frac{T}{\tau} - (1-\zeta) \kappa Z\bigg) \\
\Omega_b & \to \cos\theta\frac{2}{\tau}\text{ sech} \bigg(\frac{T}{\tau} - (1-\zeta) \kappa Z\bigg),
\end{align}
\end{subequations}
and the line-center excited state probability is now:
\begin{equation}
\label{ExcStateProbOutput}
\rho_{33} \to (1-\zeta)\text{ sech}^2\bigg(\frac{T}{\tau}-(1-\zeta)\kappa Z\bigg).
\end{equation}
The output pulses are again matched simultons, and have quite similar features to the output pulses for the pure state.  The major difference is that the dark state can no longer be fully populated so interaction between the pulses and medium continues, however because of SIT the pulses continue to propagate without absorption.  The group velocity is not $c$, rather it is $v_g/c = [1 + (1-\zeta)\kappa \tau]^{-1}$, and at the peak of the pulse the excited state probability is $\rho_{33} = 1-\zeta$ instead of 0.

\begin{figure}[h]
\begin{center}
\leavevmode
\includegraphics{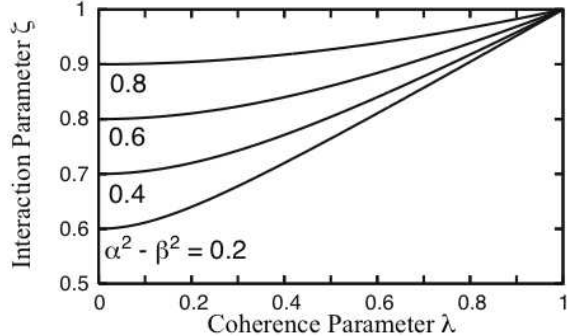}
\end{center}
\caption[Plots of the interaction parameter defined in Eq. \eqref{eigValue}.]{\label{fig.absorption-param} Plots of the interaction parameter defined in Eq. \eqref{eigValue}.  The horizontal axis is the coherence parameter $\lambda$, and the vertical axis is the interaction parameter $\zeta$. We plot the interaction parameter for a variety of media preparations ranging from $\alpha^2 -\beta^2 = 0.2$ up to $0.8$.  It ranges in value from $\zeta = \alpha^2$ for $\lambda = 0$ (assuming $\alpha^2 > \beta^2$) to $\zeta = 1$ for $\lambda = 1$.}
\end{figure}

Both the input and output regimes show modified excited state populations when compared to the pure state case, and the modification in both regimes is governed by the same parameter $\zeta$.  We will refer to this parameter $\zeta$ as the interaction parameter.  Its value gives the maximum population of the dark state, and thus determines how strongly EIT can cancel the pulse-medium interaction.  In the pure-state case $\zeta = 1$, and all interaction ceased once the pulses were in the dark state.  However as just shown, when the medium is initially in a mixed-state, we have $\zeta<1$, and interaction continues.  The value of $\zeta$ decreases until the limiting completely mixed state is reached, where $\lambda=0$, giving $\zeta = \alpha^2$ (for $\alpha^2 > \beta^2$).  In Fig. \ref{fig.absorption-param} we plot the value of this interaction parameter as a function of the coherence parameter $\lambda$ for a variety of initial medium preparations.  The plot shows the range of the parameter $\zeta$ and its dependence on initial medium preparation.  However as the medium approaches the pure-state case, all curves converge, and thus all interaction is cancelled no matter how the population is initially distributed.

\begin{figure}[h!]
\begin{center}
\leavevmode
\includegraphics[height=2.4in]{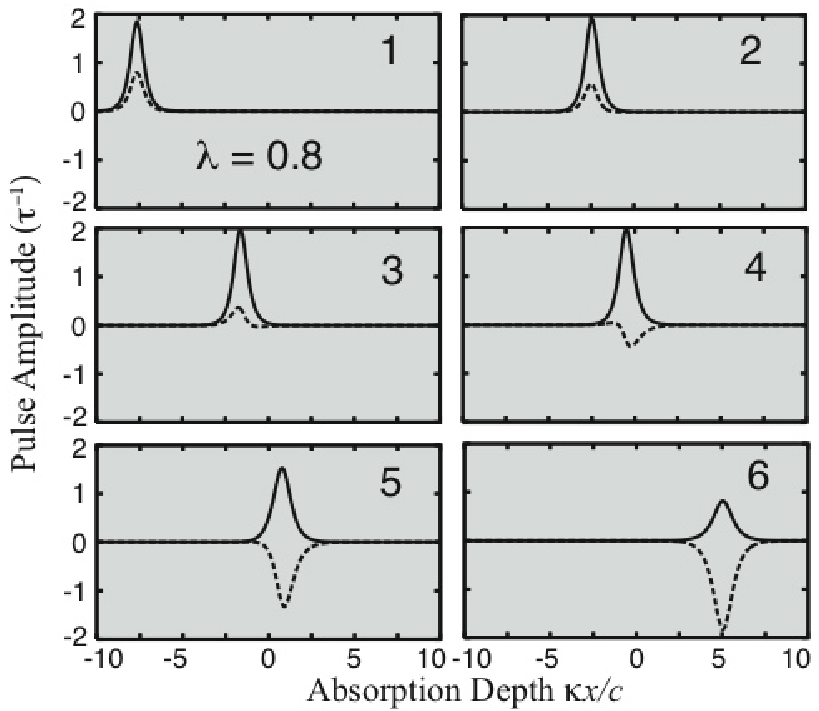}
\hspace{0.2in}
\includegraphics[height=2.4in]{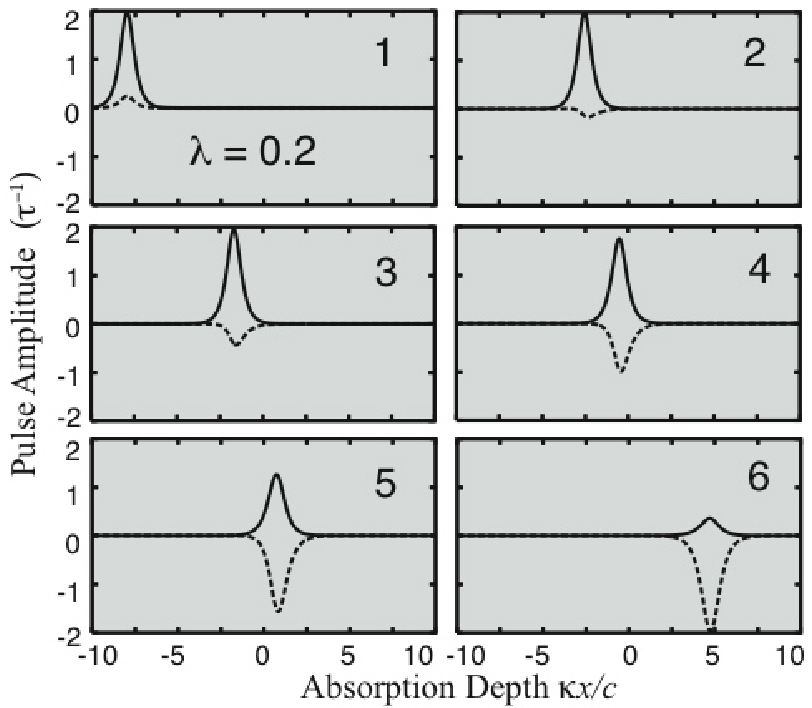}
\end{center}
\caption[Plots of the analytic pulse solutions given in Eq. \eqref{MixedStateSolutions} for a mixonium medium.]{\label{fig.mix.anPulse0.8} Plots of the analytic pulse solutions given in Eq. \eqref{MixedStateSolutions} for a mixonium medium for two different values of $\lambda$.  The horizontal axis is $x$ in units of $\kappa/c$, and the vertical axis is the pulse Rabi frequency in units of $\tau^{-1}$.  The pump pulse, $\Omega_a$, is the solid curve, and the Stokes pulse, $\Omega_b$, is the dashed curve.  The plot shows the simulton transfer process as an exchange process between bright-state simultons to dark-state simultons.  Parameters: $\alpha^2 = 0.8$, $\beta^2 = 0.2$, $\tau \approx 3T_2^*$, and $\lambda=0.8$ (left figure) $\lambda = 0.2$ (right figure).}
\end{figure}

\begin{figure}[h!]
\begin{center}
\leavevmode
\includegraphics[height=2.4in]{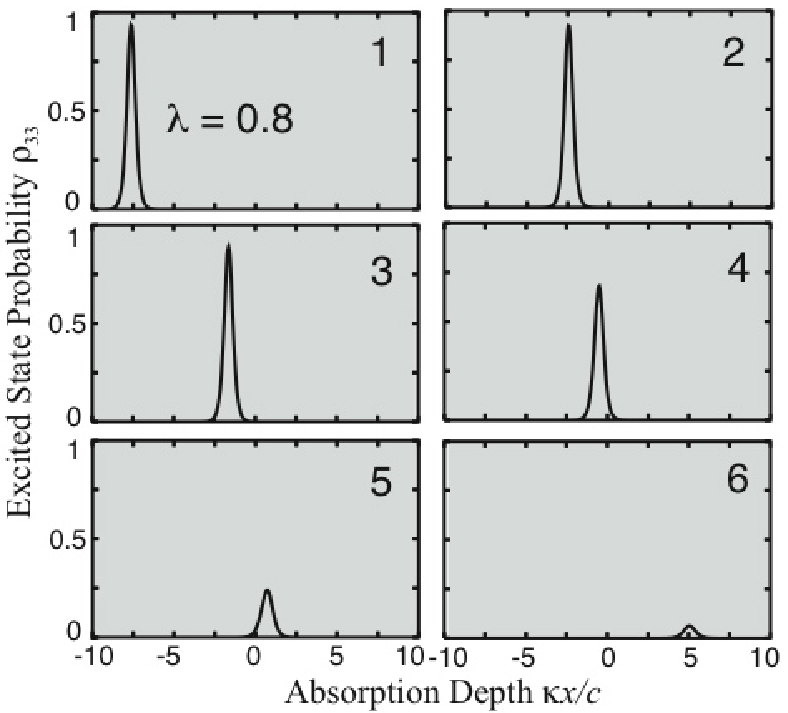}
\hspace{0.2in}
\includegraphics[height=2.4in]{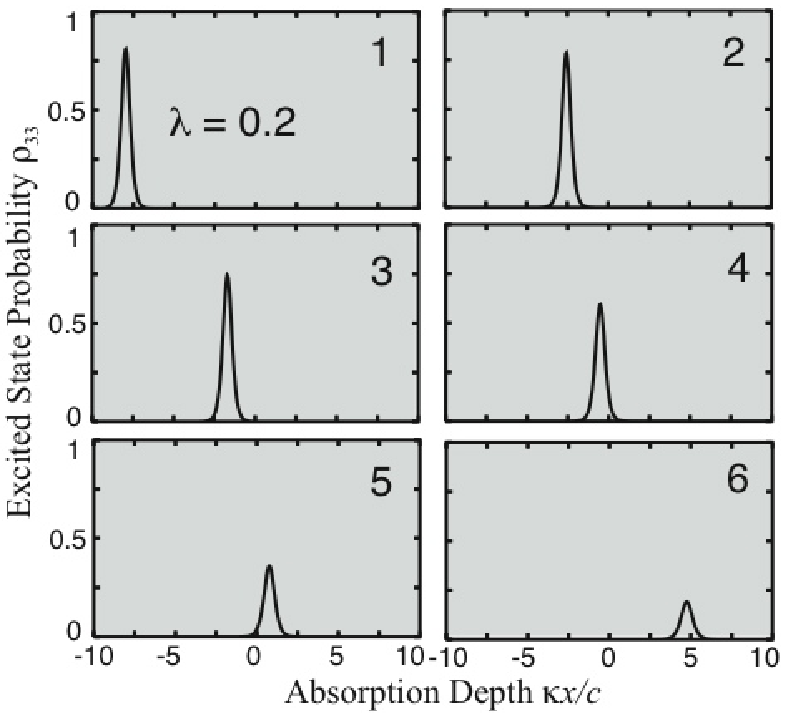}
\end{center}
\caption[Plots of the analytic excited state population solutions given in Eq. \eqref{ExStateProb} for a mixonium medium.]{\label{fig.mix.anPop0.8} Plots of the analytic excited state population solutions given in Eq. \eqref{ExStateProb} for a mixonium medium.  The horizontal axis is $x$ in units of $\kappa/c$, and the vertical axis is the excited state probability $\rho_{33}$.  The plot shows the excited state being heavily populated during the input propagation regime, when the pulses and atoms are in the bright state.  However unlike the pure-state case, the excited state never reaches $\rho_{33} = 1$.  As the pulses and atoms transfer into the dark state the pulses and atoms have a weaker interaction, however the interaction is never completely cancelled and $\rho_{33} \ne 0$.  Parameters: $\alpha^2 = 0.8$, $\beta^2 = 0.2$, $\tau \approx 3T_2^*$, and $\lambda=0.8$ (left figure) $\lambda = 0.2$ (right figure).}
\end{figure}

Next we plot the pulse solutions for a mixed-state medium with parameters $\alpha^2 - \beta^2 = 0.6$ in Fig. \ref{fig.mix.anPulse0.8}.  The left figure corresponds to $\lambda = 0.8$ and the right figure corresponds to $\lambda = 0.2$.  The solutions for the excited state population are plotted underneath in Fig. \ref{fig.mix.anPop0.8}.  For the pulses we see a very similar propagation behavior to that of the pure-state case plotted in Fig. \ref{fig.mix.anPulse1.0}.  Aside from a slight difference between the relative pulse amplitudes the plots look very alike. The main difference is seen in the plots of the excited state population.  The input pulses shown in frame 1 no longer cause complete excitation into the excited state so $\rho_{33} \ne 1$ even at the pulse peak.  More importantly the output pulses are no longer completely decoupled from the medium, since $\rho_{33} \ne 0$ in the output regime III as seen in frames 5 and 6 of Fig. \ref{fig.mix.anPop0.8}. This feature is more enhanced as the value of $\lambda$ gets smaller.  The fact that the medium is not in a fully coherent pure-state superposition implies that the dark state cannot be fully populated and thus the pulse and atoms still interact.  However because of SIT the medium is still transparent to the pulses.

%%%%%%%%%%%%%%%%%%%%%%%%%%%%%%%%%%%%%%%%%%%%%%%%%
%
%						         Pulse Area
%
%%%%%%%%%%%%%%%%%%%%%%%%%%%%%%%%%%%%%%%%%%%%%%%%%
\section{Pulse Area}
We showed in section \ref{ss:bright-dark} that in a pure-state medium, with the pulses initially matched, such that the dark state is cancelled, the medium will behave exactly as a two-level medium when described in the dressed bright-dark basis. In this basis the two pulses must be combined and thought of as one ``total" pulse, with total Rabi frequency defined as $\Omega_T = (|\Omega_a|^2 + |\Omega_b|^2)^{1/2}$.  We also found in the previous section that even though interaction occurred between the pulses and atoms in a mixonium medium, the medium still appeared transparent due to SIT.  Thus we expect some elements of the two-level area theorem to hold for the bright pulse area.  With the area of a pulse defined to be
\begin{equation}
\label{mix-pulseArea2}
A(Z) = \int_{-\infty}^{\infty}\Omega(Z,T)dT,
\end{equation}
the pulse areas of the general mixed state solutions \eqref{MixedStateSolutions} can be shown to be
\begin{subequations}
\label{mix-PulseAreas}
\begin{align}
A_{a}(Z) & = 2\pi \left(\frac{\cos\theta}{h(Z)} + \frac{\sin\theta}{h(-Z)}\right) \\
A_{b}(Z) &= 2\pi \left(\frac{-\sin\theta}{h(Z)} + \frac{\cos\theta}{h(-Z)}\right)
\end{align}
\end{subequations}
where
\begin{equation}
\label{mix-AreaFunction}
h(Z) = \sqrt{1+e^{2(2\zeta-1)\kappa Z}}.
\end{equation}

A remarkable results occurs when considering the total-pulse area:
\begin{equation}
\label{mix-totalArea}
A_T(Z) = \int_{-\infty}^{\infty}(\Omega_a^2 + \Omega_b^2)^{1/2}dT.
\end{equation}
Since the analytical solutions are temporally matched (i.e. $\partial/\partial T (\Omega_a /\Omega_b) = 0$), the total pulse area can be written as the sum of the squares of the individual pulses areas.  Thus the area of the total Rabi frequency from the analytic solutions is simply:
\begin{equation}
\label{mix-totalArea-analytic}
A_T(Z) = \sqrt{A_a^2(Z) + A_b^2(Z)} = 2\pi.
\end{equation}
This indicates a remarkably broad connection to SIT in which the pulse area must obey the area theorem:
\begin{equation}
\label{area-theorem}
\frac{1}{c}\frac{\partial A(Z)}{\partial Z} = -\frac{\alpha_D}{2}\sin A(Z).
\end{equation}
Even for an arbitrary mixed-state medium, the total pulse area remains constant and equal to $2\pi$.   We will elaborate on this SIT connection in the next section via numerical solutions.

%%%%%%%%%%%%%%%%%%%%%%%%%%%%%%%%%%%%%%%%%%%%%%%%%
%
%						         Numerical Solutions
%
%%%%%%%%%%%%%%%%%%%%%%%%%%%%%%%%%%%%%%%%%%%%%%%%%
\section{Numerical Solutions}
We now focus our attention on examining the consequences of both the modified dark state properties of a mixed state medium, and the connection of our pulse solutions to SIT.  We do this by testing our exact analytical solutions in a more realistic experimental setting, through numerical solutions to the MB equations.  This allows us to test the general utility of insights suggested by the highly specialized exact matched \textit{sech} shaped analytical input pulse solutions for infinite medium length.  We will use gaussian input pulses defined as:
\begin{equation}
\label{mix-numInput}
\Omega_{a}^{(in)}  = \frac{A_{a}}{\tau_a\sqrt{2\pi}}e^{-\frac{T^2}{2\tau_a^2}}\ \quad {\rm and}\ \quad 
\Omega_{b}^{(in)}  = \frac{A_{b}}{\tau_b\sqrt{2\pi}}e^{-\frac{T^2}{2\tau_b^2}},
\end{equation}
and ``super-gaussian" input pulses defined as:
\begin{equation}
\label{mix-numInputSquare}
\Omega_{a}^{(in)}  = \frac{A_{a}}{\tau_a\Gamma(\frac{1}{4})}e^{-\frac{T^4}{(2\tau_a)^4}}\ \quad {\rm and}\ \quad 
\Omega_{b}^{(in)}  = \frac{A_{b}}{\tau_b\Gamma(\frac{1}{4})}e^{-\frac{T^4}{(2\tau_b)^4}},
\end{equation}
where $A_a$ gives the pump pulse area as defined in \eqref{mix-pulseArea2}, $\tau_a$ is the nominal pump pulse width, $\Gamma(1/4) \approx 3.6$ is the gamma function, and similarly for the Stokes pulse.  We replace the infinite uniform medium by a medium with definite entry and exit faces.

We will first review pulse propagation in phaseonium, a pure-state medium.  We will show how the SIT-like propagation predicted in Sec. \ref{ss:bright-dark} can be realized by using matched input pulses with ratio $\Omega_a/\Omega_b = \alpha/\beta$ such that $|c_D|^2 = 0$.  However small fluctuations eventually cause these pulses to reach the dark state.  Next we will show numerical solutions with unmatched input pulses that demonstrate pulse matching, similar to results shown in ref. \cite{Kozlov-Eberly}.  We will contrast these results with solutions for pulse propagation in a mixonium medium.  Because the dark state can never be fully populated for pulses propagating in mixonium, SIT always plays a role, and EIT dominance is weakened.  The difference is most pronounced for long media that are many absorption depths long, however even short media show some of the characteristics.

%%%%%%%%%%%%%%%%%%%%%%%%%%%%%%%%%%%%%%%%%%%%%%%%%
%
%						Phaseonium - Matched Input Pulses
%
%%%%%%%%%%%%%%%%%%%%%%%%%%%%%%%%%%%%%%%%%%%%%%%%%
\subsection{Phaseonium - Matched Input Pulses}
\begin{figure}[h!]
\begin{center}
\leavevmode
\includegraphics[height=2.7in]{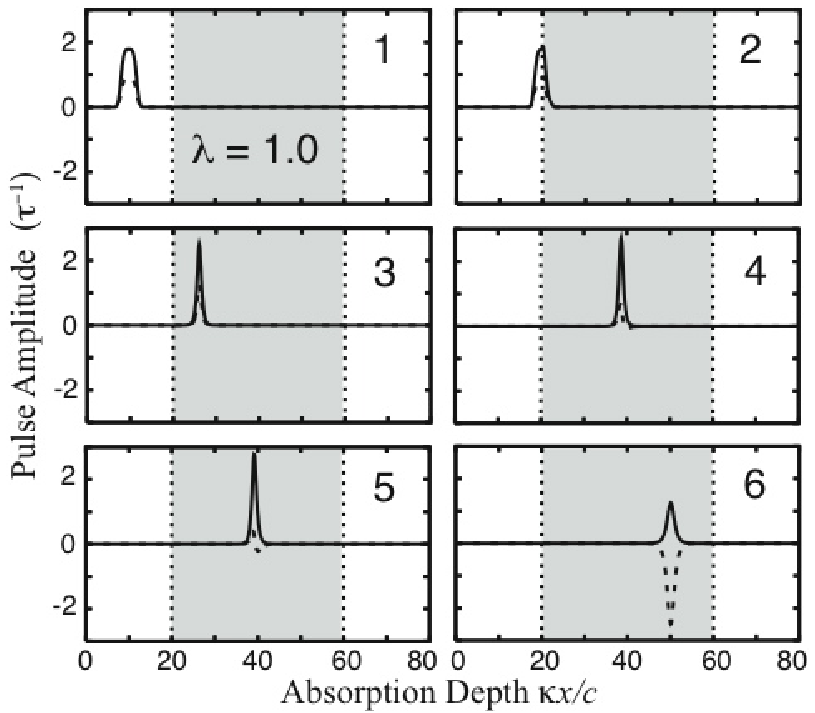}
\end{center}
\caption[Plots of numerical pulse solutions of Eqs. \eqref{rhoEquations2} for a pure-state phaseonium medium, with super-gaussian input pulses.]{\label{fig.mix.numMatchedPulse1.0} Plots of numerical pulse solutions of Eqs. \eqref{rhoEquations2} for a pure-state phaseonium medium, with super-gaussian input pulses.  The horizontal axis is $x$ in units of $\kappa/c$, and the vertical axis is the pulse Rabi frequency in units of $\tau^{-1}$. The solid curve is the pump pulse, $\Omega_{a}$, and the dashed curve is the Stokes pulse, $\Omega_{b}$.  The plot shows matched input pulses with ratio $\Omega_a/\Omega_b = \alpha/\beta$, such that $|c_D|^2 \approx 0$ reshaped just as in normal SIT.  However after some propagation distance their ratios change to $\Omega_a/\Omega_b = -\beta/\alpha$ as predicted by the dark area theorem so that $|c_D|^2 \approx 1$.  Parameters: $\alpha^2 = 0.8$, $\beta^2 = 0.2$, $\tau_a = \tau_b = \tau \approx 3T_2^*$, $A_a = 2.3\alpha \pi$, $A_b = 2.3\beta\pi$, and $\lambda=1.0$.  The value of $\mu$ is chosen so that $v_g/c = 1/2$ inside the medium.}
\end{figure}

As shown in Sec. \ref{ss:bright-dark}, when the pulses are initially matched, with ratios $\Omega_a/\Omega_b = \alpha/\beta$, such that $|c_D|^2 = 0$, the three-level $\Lambda$ system reduces to that of a two-level system regardless of the composite bright-pulse shape.  We plot such a solution in Fig. \ref{fig.mix.numMatchedPulse1.0}, where we use matched super-gaussian input pulses as defined in Eq. \eqref{mix-numInputSquare}, with equal pulse widths $\tau_a = \tau_b = 3T_2^*$ and areas $A_a = 2.3 \alpha \pi$ and $A_b = 2.3 \beta \pi$.  The medium is prepared in a pure state such that $\lambda = 1.0$ with populations $\alpha^2 - \beta^2 = 0.6$.  This solution clearly illustrates the three regime language that we have introduced.  In regime I (frames 1-3) the pulses act as simultons, and the pulse-medium interaction is exactly that for resonant two-level pulse propagation when considered in the dressed state basis.  Then in regime II (frames 4-5) the pulse amplitudes rapidly change leading to the dark-state output regime III (frame 6).  The solution exhibits SIT type propagation in regime I, where $|c_D|^2 \approx 0$, but beginning near frame 4 the pulse amplitudes begin to change, and eventually end up with ratio $\Omega_a/\Omega_b = -\beta/\alpha$ in frame 6.  Here all population is in the dark-state and $|c_D|^2 \approx 1$, and EIT type propagation takes over.  

\begin{figure}[h!]
\begin{center}
\leavevmode
\includegraphics[height=2.0in]{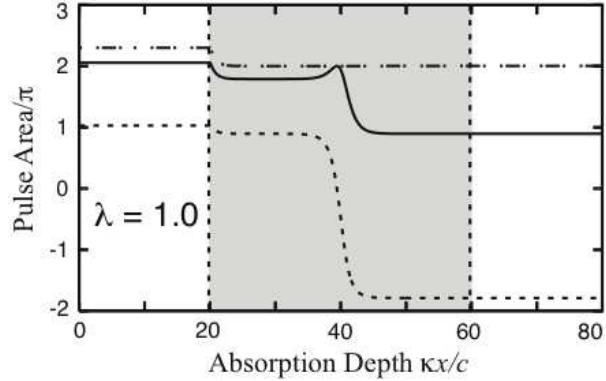}
\end{center}
\caption[Numerically integrated area of the individual pulse areas as well as the total Rabi frequency area for the pulse solutions shown in Fig. \ref{fig.mix.numMatchedPulse1.0}.]{\label{fig.mix.MatchedArea1.0} Numerically integrated area of the individual pulse areas as well as the total Rabi frequency area for the pulse solutions shown in Fig. \ref{fig.mix.numMatchedPulse1.0}.  The vertical axis is the pulse area. The solid curve is the area of the pump pulse, the dashed curve is the area of the Stokes pulse, and the dot-dashed curve is the area of the total Rabi frequency.  The medium initially behaves as a two-level medium, and the bright area changes to $2\pi$.  However after some propagation distance the pulses transfer to the dark state, while the bright area remains constant through this change.}
\end{figure}

The initial SIT type behavior is further illustrated in Fig. \ref{fig.mix.MatchedArea1.0} where we plot the areas of the individual pulses, as well as the area of the bright pulse.  After a few absorption depths, the total pulse area is quickly changed from its input area of $A_T = 2.3\pi$ to $2\pi$.  Then after propagating as simultons for a while, the pulses quickly change to the dark state.  While the individual pulse areas are rapidly changing, we see completely static behavior for the bright area, confirming our analytic result in Eq. \eqref{mix-totalArea}.

After the initial reshaping caused by the two-level SIT behavior, the pulses propagate for some time as simultons.  The simulton solutions are exact solutions, so the reason for the pulses to transfer to the dark-state does not readily present itself.  Only perfect matched {\it sech} shaped pulses remain as simultons.  Any small perturbation from these exact shapes implies $|c_D|^2 > 0$.  While the initial dark-state probability may be small, any non-zero value will eventually lead to the dark state if the propagation distance is long enough.  In this particular example, the small perturbations are caused by numerical roundoff error. The analytic solutions actually predict this behavior.  It is the exponential term in the denominator of Eqs. \eqref{DiagPulseSol} which describes deviations from the simulton shape.  Thus we see that simulton pulse propagation is in fact an inherently unstable propagation scenario, and that small fluctuations will always lead to the dark state.

%%%%%%%%%%%%%%%%%%%%%%%%%%%%%%%%%%%%%%%%%%%%%%%%%
%
%						       Phaseonium - Mismatched Input Pulses
%
%%%%%%%%%%%%%%%%%%%%%%%%%%%%%%%%%%%%%%%%%%%%%%%%%
\subsection{Phaseonium - Mismatched Input Pulses}
\begin{figure}[h!]
\begin{center}
\leavevmode
\includegraphics[height=2.9in]{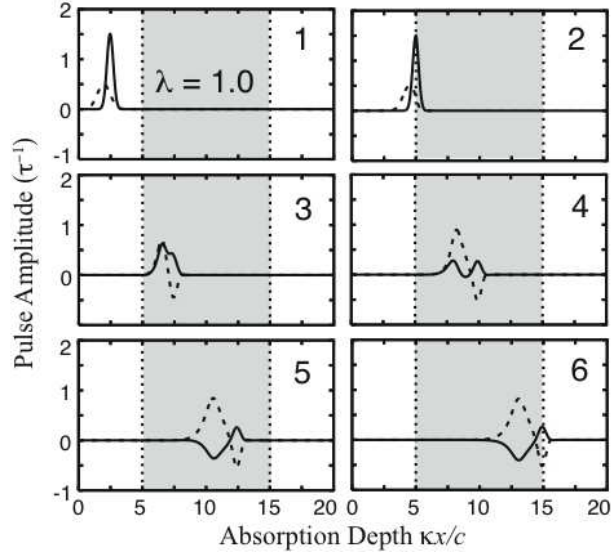}
\end{center}
\caption[Plots of numerical pulse solutions of Eqs. \eqref{rhoEquations2} for a pure-state phaseonium medium, with gaussian input pulses.]{\label{fig.mix.numPulse1.0Short} Plots of numerical pulse solutions of Eqs. \eqref{rhoEquations2} for a pure-state phaseonium medium, with gaussian input pulses.  The horizontal axis is $x$ in units of $\kappa/c$, and the vertical axis is the pulse Rabi frequency in units of $\tau^{-1}$. The solid curve is the pump pulse, $\Omega_{a}$, and the dashed curve is the Stokes pulse, $\Omega_{b}$.  The plot shows mis-matched input pulses quickly reshaped into matched pulses with ratios $\Omega_a/\Omega_b = -\beta/\alpha$ as predicted by the dark area theorem and the analytic solutions.  Parameters: $\alpha^2 = 0.8$, $\beta^2 = 0.2$, $\tau_a = \tau_b/2 = \tau \approx 3T_2^*$, temporal mismatch of $2\tau_a$, $A_a = 1.2\pi$, $A_b=0.8\pi$, and $\lambda=1.0$.}
\end{figure}

Next we examine what happens if the pulses are initially temporally mismatched, but where the medium is still prepared in a pure state.  We show the numerical pulse solutions in Fig. \ref{fig.mix.numPulse1.0Short}, where we take the medium to be prepared with $\alpha^2 - \beta^2 =0.6$, and the pulses to be gaussian shape with width $\tau_a = \tau_b/2 = 3T_2^*$, a temporal mismatch of $2\tau_a$, and areas $A_a = 1.2\pi$ and $A_b = 0.8\pi$.  In this case, the pulses do not propagate as simulton pulses, since they are not initially matched (i.e. $\Omega_a/\Omega_b \ne \text{ constant}$).  However as they propagate they are quickly reshaped into two-peaked but matched pulses with ratio $\Omega_a/\Omega_b = -\beta/\alpha$, such that the dark-state is fully populated and EIT type propagation occurs.  This re-shaping occurs because the dark Rabi frequency is nonzero for mismatched pulses.  In this example the medium is only 10 absorption depths long, and most re-shaping is complete after about 5 absorption depths.  This plot confirms previous KE results showing very similar behavior \cite{Kozlov-Eberly}.

\begin{figure}[h!]
\begin{center}
\leavevmode
\includegraphics[height=2.1in]{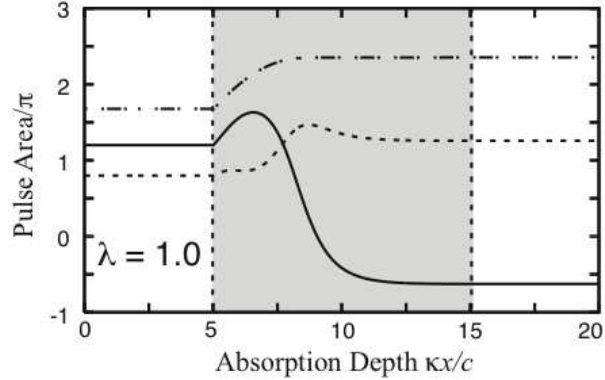}
\end{center}
\caption[Numerically integrated area of the individual pulse areas as well as the total Rabi frequency area for the pulse solutions shown in Fig. \ref{fig.mix.numPulse1.0Short}.]{\label{fig.mix.Area1.0Short} Numerically integrated area of the individual pulse areas as well as the total area for the pulse solutions shown in Fig. \ref{fig.mix.numPulse1.0Short}.  The horizontal axis is $x$ in units of $\kappa/c$, and the vertical axis is the pulse area. The solid curve is the area of the pump pulse, the dashed curve is the area of the Stokes pulse, and the dot-dashed curve is the area of the total Rabi frequency.  The pulses initially reshape as they match, however after a short propagation distance the pulses enter the dark state, and the pulse areas become constant.  The SIT area theorem does not apply in this case.}
\end{figure}

We also plot numerically integrated pulse areas in Fig. \ref{fig.mix.Area1.0Short} for both the individual pulse areas as well as the area of the total Rabi frequency.  This plot further illustrates the distinction between the matched and mismatched cases.  The pulse areas are quickly modified as the pulses match, but as soon as they reach the dark state, the pulse areas become constant.  The SIT area theorem does not apply in this case, and no prediction is possible as to the final pulse areas, unlike the previous example.

These two examples of matched and mis-matched input pulses highlight a feature common to pulse propagation in phaseonium.  That is, the dark-state always dominates.  Pulses will always end up matched and always with a ratio that satisfies the dark area theorem, so that absorption is cancelled and EIT plays a dominate role.  However, for matched input pulses with no dark state population, two-level physics initially describes the propagation, and the dark-state dominance takes much longer to appear.    While we can identify the input, output and transfer regimes in the matched input example, we cannot do the same for the mismatched input pulses.  We will see in the next section that the mixonium medium modifies the absorptive properties causing the dark-state dominance to be replaced with SIT like effects.

%%%%%%%%%%%%%%%%%%%%%%%%%%%%%%%%%%%%%%%%%%%%%%%%%
%
%						Mixonium - Mismatched Input Pulses
%
%%%%%%%%%%%%%%%%%%%%%%%%%%%%%%%%%%%%%%%%%%%%%%%%%
\subsection{Mixonium - Mismatched Input Pulses}
We now examine the effects that mixonium has on mismatched pulse propagation.  Matched input pulses (no matter the shape) with input ratios given by the analytic solutions in Eqs. \eqref{MixedPulseInput} will behave in a similar manner to the matched pulse example in the previous section, so we do not plot the results here.  The difference is simply that the dark state can no longer be fully populated as discussed in Sec. \ref{ss:mixed-state-an}, and thus the pulses continue to cause excitation into the excited state.

\begin{figure}[h!]
\begin{center}
\leavevmode
\includegraphics[height=2.8in]{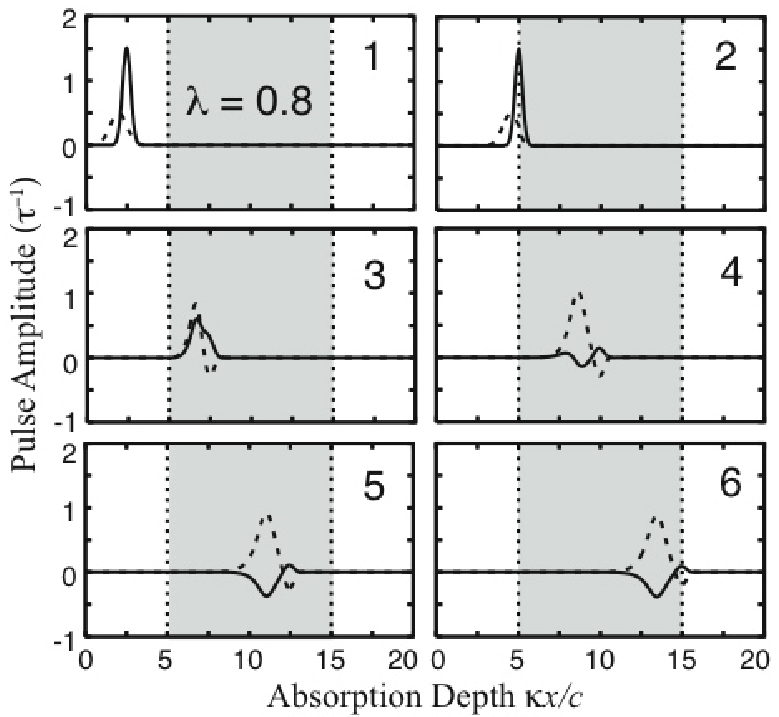}
\hspace{0.1in}
\includegraphics[height=2.8in]{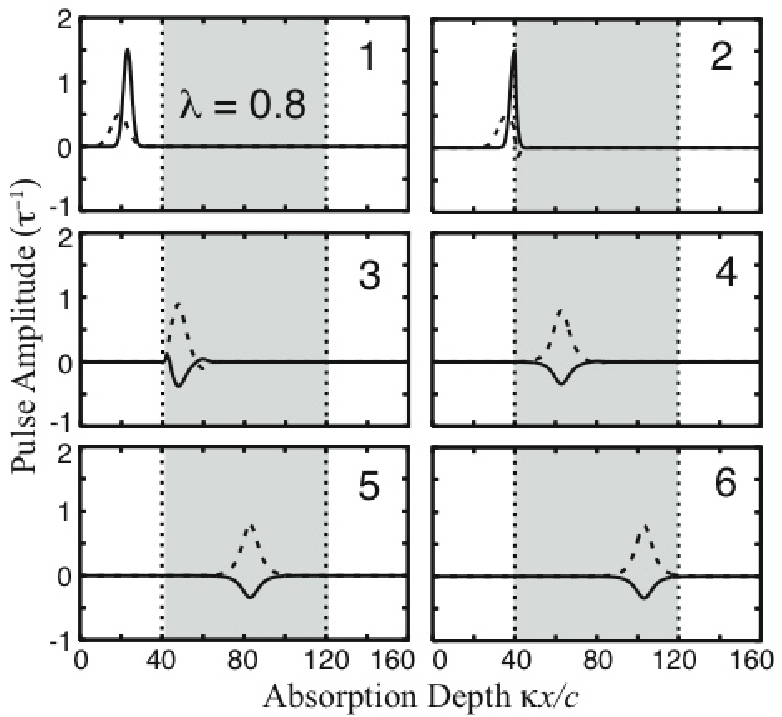}
\end{center}
\caption[Plots of numerical pulse solutions of Eqs. \eqref{rhoEquations2} for a mixed-state medium, with gaussian input pulses.]{\label{fig.mix.numPulse0.8Short}\label{fig.mix.numPulse0.8Long} Plots of numerical pulse solutions of Eqs. \eqref{rhoEquations2} for a mixed-state medium, with gaussian input pulses.  The horizontal axis is $x$ in units of $\kappa/c$, and the vertical axis is the pulse Rabi frequency in units of $\tau^{-1}$.  The left and right frames are identical except for different medium lengths.  The solid curve is the pump pulse, $\Omega_{a}$, and the dashed curve is the Stokes pulse, $\Omega_{b}$.  The plot shows mis-matched input pulses quickly reshaped into matched pulses, similar to the pure state case.  Parameters: $\alpha^2 = 0.8$, $\beta^2 = 0.2$, $\tau_a = \tau_b/2 = \tau \approx 3T_2^*$, temporal mismatch of $2\tau_a$, $A_a = 1.2\pi$, $A_b=0.8\pi$, and $\lambda=0.8$.}
\end{figure}

\begin{figure}[h!]
\begin{center}
\leavevmode
\includegraphics[height=1.9in]{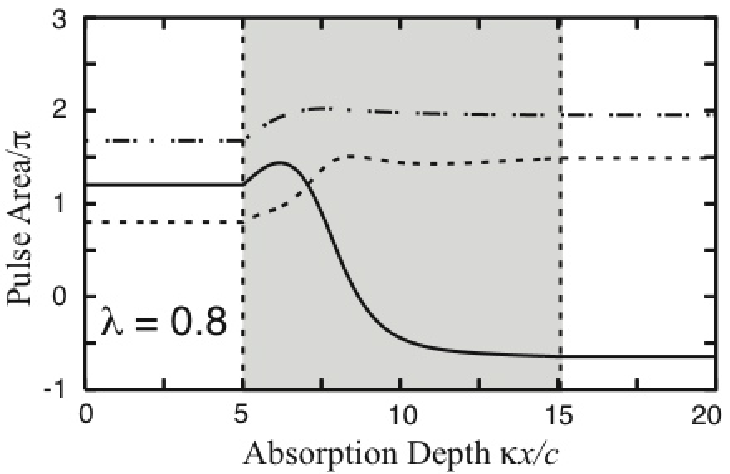}
\hspace{0.1in}
\includegraphics[height=1.9in]{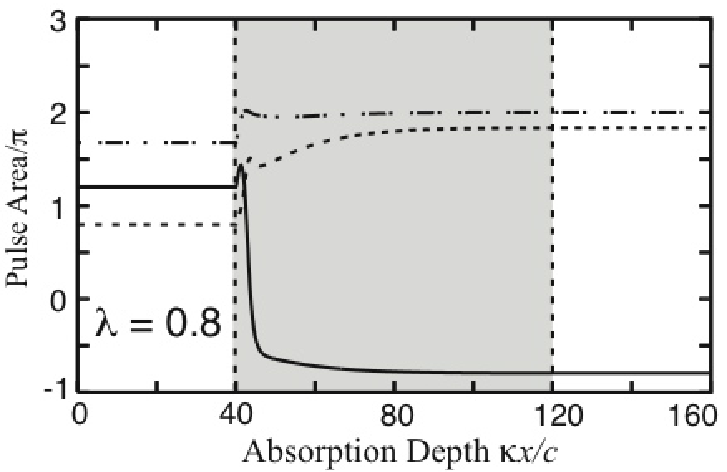}
\end{center}
\caption[Numerically integrated area of the individual pulse areas as well as the total Rabi frequency area for the pulse solutions shown in Fig. \ref{fig.mix.numPulse0.8Long}.]{\label{fig.mix.Area0.8Short} \label{fig.mix.Area0.8Long}Numerically integrated area of the individual pulse areas as well as the total Rabi frequency area for the pulse solutions shown in Fig. \ref{fig.mix.numPulse0.8Long}.  The horizontal axis is $x$ in units of $\kappa/c$, and the vertical axis is the pulse area. The solid curve is the area of the pump pulse, the dashed curve is the area of the Stokes pulse, and the dot-dashed curve is the area of the total Rabi frequency.  We see initially rapid change in the pulse areas as they are matched, consistent with pure-state behavior.  However because the dark state can never be fully populated the medium always behaves as an SIT like medium, and eventually the bright pulse area changes to $2\pi$ area.}
\end{figure}

The inability of the dark state to be fully populated in mixonium has a profound effect on mis-matched pulses propagating through many absorption depths (the result is less dramatic for short media). We plot the pulse solutions for the same parameters as the solutions plotted in the previous section, with $\alpha^2 - \beta^2 = 0.6$, gaussian pulse shapes with duration $\tau_a = \tau_b/2 = 3T_2^*$ and an offset of $2\tau_a$, and pulse areas of $A_a = 1.2\pi$ and $A_b = 0.8\pi$, except we take the medium to be in a mixed state with $\lambda = 0.8$.  We plot these pulse solutions in the left frame of Fig. \ref{fig.mix.numPulse0.8Short}.  We see a very similar behavior to the previous solutions with one almost unnoticeable difference. That is, the output pulse ratio is given by $\Omega_a/\Omega_b = \tan \theta = -\lambda\alpha\beta/(\zeta - \beta^2)$ as predicted by the mixed-state analytic solutions.  However, when we plot the areas of the pulses in the left frame of Fig. \ref{fig.mix.Area0.8Short} we notice now that the total pulse area is very close to $2\pi$, in contrast to the pure-state solution shown in Fig. \ref{fig.mix.Area1.0Short}.  In fact in this example, the pulses were still being reshaped, thus we will examine what happens if the medium is slightly longer.

\begin{figure}[h!]
\begin{center}
\leavevmode
\includegraphics[height=2.6in]{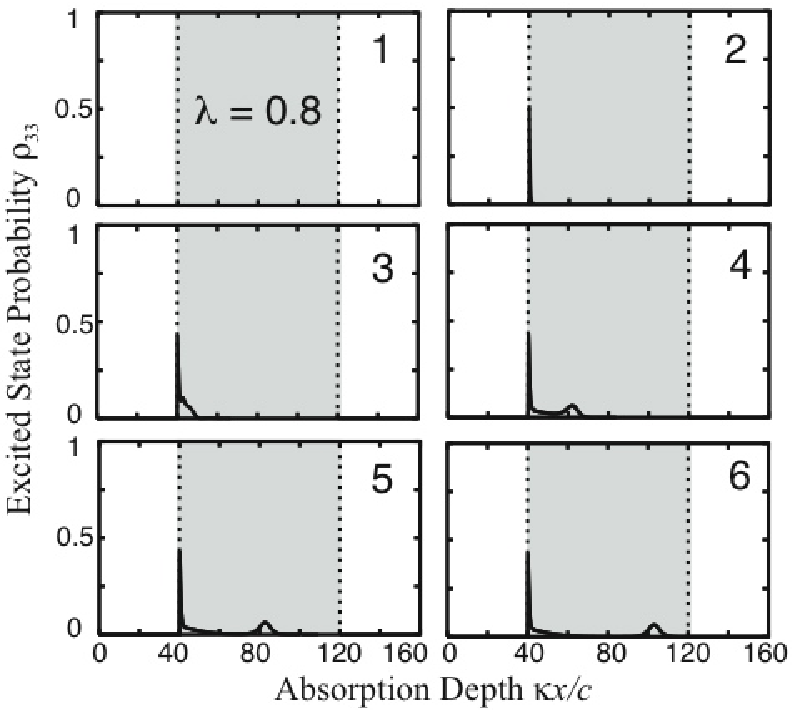}
\end{center}
\caption[Plots of numerical excited state population solutions of Eqs. \eqref{rhoEquations2} for a mixed-state medium.]{\label{fig.mix.numPop0.8Long} Plots of numerical excited state population solutions of Eqs. \eqref{rhoEquations2} for a mixed-state medium.  Each frame corresponds exactly to the same frame on the right hand side of Fig. \ref{fig.mix.numPulse0.8Long}.   The horizontal axis is $x$ in units of $\kappa/c$, and the vertical axis is the excited state population. The plot shows that even after the pulses are matched, the dark state can never be fully populated in a mixed-state medium, and thus the excited state continues to be populated.  Parameters: $\alpha^2 = 0.8$, $\beta^2 = 0.2$, $\tau_a = \tau_b/2 = \tau \approx 3T_2^*$, temporal mismatch of $2\tau_a$, $A_a = 1.2\pi$, $A_b=0.8\pi$, and $\lambda=0.8$.}
\end{figure}

The difference between phaseonium and mixonium propagation is greatly magnified when the pulses propagate through a longer medium.  We plot these solutions in the right frame of Fig. \ref{fig.mix.numPulse0.8Long} which show continued reshaping of the pulses until they are matched {\it sech} shaped pulses, exactly as given by the output analytic solutions in Eqs. \eqref{MixedPulseOutput}.  We plot the pulse areas for this example in the right frame of Fig. \ref{fig.mix.Area0.8Long} and we see that after the initial rapid reshaping which matches the pulses, the areas of the pulses continue to change until the bright pulse area reaches $A_b = 2\pi$.  The modified interaction properties of mixonium cause SIT to dominate and reshape the pulses.  Thus the EIT dominance that was exhibited in the pure-state case, is now replaced with SIT dominance for mixed-state media.  For this same long medium example we also plot the excited state population in Fig. \ref{fig.mix.numPop0.8Long}.  We see that even after the pulses are matched, the excited state can never be completely decoupled.  Thus, the dark state can never be fully populated, with its maximum value given by the interaction parameter $\zeta$.  The inability of the dark state to be fully populated allows SIT to continuously reshape the pulses until they agree with the analytic solutions.

%%%%%%%%%%%%%%%%%%%%%%%%%%%%%%%%%%%%%%%%%%%%%%%%%
%
%						      Conclusions
%
%%%%%%%%%%%%%%%%%%%%%%%%%%%%%%%%%%%%%%%%%%%%%%%%%
\section{Conclusions}\label{ss:mix-conclusions}
We have presented new solutions of the Maxwell-Bloch equations, both analytic and numerical, applicable to a ``mixonium" medium, where the term mixonium implies a $\Lambda$ medium prepared in a partially phase-coherent superposition of the ground states.  This medium offers a new contrast to a pure ``phaseonium" medium, where the ground states are prepared in a completely phase-coherent superposition of the two ground states.  The partially coherent medium is experimentally realistic, whereas pure-state preparations are difficult to achieve.

The analytic solutions for the pulses and density matrix elements were obtained via the Park-Shin B\"acklund transformation method \cite{park-shin}.  Consistent with our previous work \cite{clader-eberly07, clader-eberly-pra07}, we again identified three distinct regimes of interest for the analytic solutions.  For the pure-state case, we identified our solutions in the asymptotic input regime to be equivalent to the well known simulton solutions \cite{Konopnicki-Eberly}.  Our analytic solutions then describe the transfer of these input simulton pulses to simulton solutions completely in the dark state in the output regime.  We identify this behavior as the transfer of pulses propagating with completely SIT like behavior to completely EIT like behavior, where we use the same definition of SIT and EIT like behavior as given in Ref. \cite{Kozlov-Eberly}.  Using numerical solutions to the Maxwell-Bloch equations we were able to show this SIT simulton to EIT simulton behavior for pulses with matched but different shapes from the analytic solutions.  

The analytic solutions in the general mixed-state case, allow us to identify an interaction parameter that determines the maximum population of the dark state.  In the pure-state case, the dark state can become fully populated and all interaction between the pulses and medium is eliminated.  However in the mixed-state case the dark state can never be fully populated and thus the excited state cannot be decoupled.  We studied the effects that this modified interaction parameter has on dark-state propagation dynamics by numerically solving the Maxwell-Bloch equations.  In the pure-state case,  our numerical results confirmed previous results that mismatched input pulses with gaussian shapes quickly match and propagate unchanged as the dark-state is fully populated prohibiting any coupling to the excited state \cite{Kozlov-Eberly}.  In contrast, in the mixed-state case these same mismatched pulses never reach complete EIT type propagation since the dark state is never fully populated.  Unlike the pure-state case, EIT type effects cannot completely cancel SIT type effects, and the pulses continue to be reshaped into matched {\it sech} shaped simultons, matching shape to our output analytic solutions.

We have been able to demonstrate that the two-level McCall-Hahn area theorem still plays a role even in this three-level system.  The composite or ``total" two-pulse Rabi frequency of the analytic solutions has constant $2\pi$ area during all stages of propagation.  SIT propagation effects cannot be completely cancelled for mixed-state propagation, causing the composite total Rabi frequency area to evolve toward $2\pi$ just as in single pulse two-level SIT.  Only in the pure-state case where EIT can completely cancel SIT, can this behavior be avoided.  Thus we see that the two-level area theorem continues to play a role in two-pulse propagation through a three-level medium suggesting a possible three-level area theorem for the total Rabi frequency.

\acknowledgments
We thank Q-Han Park for helpful discussions and correspondence.  B.D. Clader acknowledges receipt of a Frank Horton Fellowship from the Laboratory for Laser Energetics, University of Rochester. Research has been supported by NSF Grant PHY 0456952 and PHY 0601804.  The e-mail contact address is: dclader@pas.rochester.edu.

%\bibliography{bibliography}

\end{document}